\documentclass[a4paper,fleqn]{cas-sc}

\usepackage[numbers]{natbib}
\usepackage{makecell}
\usepackage{booktabs}
\usepackage{color, colortbl}
\usepackage{array}
\usepackage{multirow}
\usepackage{nccmath}
\usepackage{appendix}

\definecolor{Gray}{gray}{0.9}
\newcommand{\ie}{\emph{i.e.},~}
\newcommand{\eg}{\emph{e.g.},~}
\newcommand{\ea}{\emph{et al.}~}
\newcommand{\vs}{\emph{vs.}~}

\def\tsc#1{\csdef{#1}{\textsc{\lowercase{#1}}\xspace}}
\tsc{WGM}
\tsc{QE}

\begin{document}
\let\WriteBookmarks\relax
\def\floatpagepagefraction{1}
\def\textpagefraction{.001}

\shorttitle{Deep grading for MRI-based differential diagnosis of Alzheimer's disease and Frontotemporal dementia}    

\shortauthors{HD Nguyen}  

\title [mode = title]{Deep grading for MRI-based differential diagnosis of Alzheimer's disease and Frontotemporal dementia}

%

\author[1]{Huy-Dung Nguyen}[orcid=0000-0002-3980-8029]
\cormark[1]

\ead{huy-dung.nguyen@u-bordeaux.fr}

\author[1]{Michaël Clément}

\author[2, 3]{Vincent Planche}

\author[1]{Boris Mansencal}

\author[1]{Pierrick Coupé}

\affiliation[1]{organization={Univ. Bordeaux, CNRS, Bordeaux INP, LaBRI, UMR 5800},
    postcode={33400 Talence},
    country={France}}

\affiliation[2]{organization={Univ. Bordeaux, CNRS, Institut des Maladies Neurodégénératives, UMR 5293},
    postcode={33000 Bordeaux},
    country={France}}

\affiliation[3]{organization={Centre Mémoire Ressources Recherches, Pôle de Neurosciences Cliniques, CHU de Bordeaux},
    postcode={33000 Bordeaux},
    country={France}}

\cortext[1]{Corresponding author}

\begin{abstract}
Alzheimer's disease and Frontotemporal dementia are common forms of neurodegenerative dementia. Behavioral alterations and cognitive impairments are found in the clinical courses of both diseases, and their differential diagnosis can sometimes pose challenges for physicians. Therefore, an accurate tool dedicated to this diagnostic challenge can be valuable in clinical practice. However, current structural imaging methods mainly focus on the detection of each disease but rarely on their differential diagnosis. In this paper, we propose a deep learning-based approach for both disease detection and differential diagnosis. We suggest utilizing two types of biomarkers for this application: structure grading and structure atrophy. First, we propose to train a large ensemble of 3D U-Nets to locally determine the anatomical patterns of healthy people, patients with Alzheimer's disease and patients with Frontotemporal dementia using structural MRI as input. The output of the ensemble is a 2-channel disease's coordinate map, which can be transformed into a 3D grading map that is easily interpretable for clinicians. This 2-channel disease's coordinate map is coupled with a multi-layer perceptron classifier for different classification tasks. Second, we propose to combine our deep learning framework with a traditional machine learning strategy based on volume to improve the model discriminative capacity and robustness. After both cross-validation and external validation, our experiments, based on 3319 MRIs, demonstrated that our method produces competitive results compared to state-of-the-art methods for both disease detection and differential diagnosis.
\end{abstract}




\maketitle

\section{Introduction}
Alzheimer’s disease (AD) and Frontotemporal dementia (FTD) are the two most common neurodegenerative causes leading to cognitive impairment and dementia \cite{bang_frontotemporal_2015}. AD is more common than FTD for people over 65, but in the 45 to 65 age range, FTD is almost as common as AD. There are some differences between the two diseases. AD patients have more problems with visuospatial abilities, while FTD patients have more frequent and severe behavioral changes  \footnote{\url{https://www.alz.org/alzheimers-dementia/what-is-dementia/types-of-dementia/frontotemporal-dementia}}. However, there are also a lot of overlapping symptoms, such as episodic memory loss, dysexecutive syndrome and/or language impairment ~\cite{boeve_advances_2022}. Accurate differential diagnosis is essential for the management of a patient’s daily life and for the implementation of dedicated clinical trials. However, the similar symptoms mentioned above make the diagnosis challenging, although the two diseases have different clinical diagnostic criteria \cite{rascovsky_sensitivity_2011, mckhann_diagnosis_2011}. Moreover, the prevalence of FTD is lower compared to AD (about 300-fold smaller) \cite{duara_frontotemporal_1999}, limiting our knowledge about FTD. Indeed, many studies have demonstrated that isolated cognitive tests cannot reliably distinguish FTD from AD populations \cite{hutchinson_neuropsychological_2007, yew_lost_2013}.
Consequently, an accurate differential diagnosis method would be beneficial for patients, families, and caregivers. In particular, a multi-class differential diagnostic tool that could distinguish between AD, FTD, and cognitively normal (CN) people would be extremely helpful in clinical practice. Indeed, such a tool can help clinicians review their hypotheses, thus making more informed decisions.

Several studies have demonstrated that AD and FTD can be individually detected using structural magnetic resonance imaging (sMRI) \cite{du_different_2006, moller_alzheimer_2016}. The areas of atrophy caused by the two diseases may differ \cite{davatzikos_individual_2008}. For instance, AD seems to mainly affect the medial temporal area \cite{jack_medial_1997} while FTD affects different regions depending on its sub-types \cite{lu_patterns_2013}. The behavioral variant frontotemporal dementia (bvFTD) is often associated with atrophy in the frontal and anterior temporal region. Patients with Progressive non-fluent aphasia (PNFA) have motor speech impairments, mainly controlled by the left inferior frontal lobe. The semantic variant (SV) mainly affects the left anterior temporal area \cite{Planche_2023_anatomical}. Hence, using sMRI for disease classification and differential diagnosis should be beneficial. Indeed, some approaches have previously been proposed to address these problems using volumetric and shape measurements extracted from sMRI \cite{du_different_2006, rabinovici_distinct_2008}. However, most existing methods focus only on binary classification tasks (\ie AD \vs CN, FTD \vs CN and AD \vs FTD). While the multi-class diagnosis provides potential value in clinical practice, only a few studies consider this problem \cite{bron_multiparametric_2017, kim_machine_2019, ma_differential_2020, hu_deep_2021}. Additionally, current approaches mainly use traditional machine learning techniques with handcrafted features that might not fully include all disease patterns. As a result, deep learning techniques have lately been explored. However, the outcomes of these methods are usually difficult to understand. This limitation hinders our understanding of these neurodegenerative diseases.

Recently, we proposed an interpretable framework called Deep Grading \cite{nguyen_deep_2021} for differential diagnosis between CN, AD, and FTD \cite{nguyen_2022}. In this approach, we employed a large number of U-Nets (125 models) to analyze different brain locations and generate a 3D grading map that estimates the brain abnormality level at the voxel level. This grading map was then used to compute averaged grading scores for 133 brain structures, which were subsequently fed into a Graph Convolutional Network \cite{kipf_semi-supervised_2017} for classification. The advantage of the method is that the 3D interpretable grading map can help to visualize the disease-related regions. However, this framework can only determine whether a brain region exhibits abnormality without specifying the specific disease associated with that abnormality. Furthermore, we solely consider one syndromic presentation of FTD (\ie behavioral variant) in that approach. Consequently, our understanding of the differences between AD and FTD still presents some limitations.

In this paper, we propose a method for both disease detection (\ie AD + FTD \vs CN, AD \vs CN, FTD \vs CN) and differential diagnosis (\ie AD \vs FTD and CN \vs AD \vs FTD). Our purpose is to expand our knowledge about different dementia types and provide an accurate tool for a real clinical scenario. To this end, our contributions are two-fold. Firstly, we extend the Deep Grading (DG) framework \cite{nguyen_deep_2021} by introducing multi-channel Disease's Coordinate (DC) maps. These maps enable the detection of specific disease-related patterns (\eg AD-like or FTD-like patterns) in different brain regions. Unlike considering AD and FTD as a single class as in \cite{nguyen_2022}, our DC maps allow for differentiation between AD and FTD patterns. Furthermore, these maps can be transformed into 3D interpretable grading maps, with distinct colors representing CN, AD, and FTD, facilitating clinicians in gaining deeper insights into AD and FTD pathologies. Additionally, the DC map can be coupled with a multi-layer perceptron (MLP) for classification. Secondly, we propose an ensemble approach that combines the decision of our MLP with a support vector machine (SVM) using brain structure volumes. This combination improves the model's classification performance and enhances its generalization capacity. By leveraging both the structure grading and structure atrophy information, our proposed framework demonstrates state-of-the-art performance in disease detection and differential diagnosis tasks.

This paper is an extension of the conference paper \cite{nguyen_2022}, with (i) a multi-channel extension of the DG framework capable of separating AD-like patterns and FTD-like patterns, (ii) a comparison with state-of-the-art methods using the same data for training and testing and (iii) an interpretation of the grading map for different sub-types of FTD.
\section{Materials}

\subsection{Datasets}
The data used in this study includes 3319 MRIs selected at the baseline from multiple open access databases: the Alzheimer’s Disease Neuroimaging Initiative (ADNI2) \cite{jack_alzheimers_2008}, the Frontotemporal lobar Degeneration Neuroimaging Initiative (NIFD) \footnote{Available at \url{https://ida.loni.usc.edu/}} and the National Alzheimer’s Coordinating Center (NACC) \cite{beekly_national_2007}. As the majority of MRIs with FTD are acquired with 3 Tesla machines, only 3T MRIs are selected for each class. The purpose of this is to avoid possible bias due to the acquisition protocol of different databases \cite{thibeau-sutre_mri_2022}. We use ADNI2 (\ie 180 CN and 149 AD) and NIFD (\ie 136 CN and 150 FTD) to perform a 10-fold cross-validation. We apply the stratified split strategy to alleviate the bias due to the imbalanced nature of different available classes. The cross-validation result is denoted as in-domain performance. We additionally evaluate our framework on an external dataset (\ie NACC with visits conducted between September 2005 and November 2021) to assess the generalization capacity of the compared methods or out-of-domain performance. Table \ref{tbl1} summarizes the demographic of the subjects used in this study. We only use the three sub-types of FTD in NIFD dataset: bvFTD, PNFA and SV. The reason for this is that the other variant of FTD (\ie logopenic variant) is typically associated with AD neuropathological changes \cite{henry_logopenic_2010, beber_logopenic_2014}. Finally, only subjects with consistent diagnosis thorough their follow-up sessions are included in this study.

\begin{table}[htbp]
\caption{Summary of participants used in our study. Data used for training are in bold, therefore MRIs from ADNI2 and NIFD are in-domain data while MRIs from NACC dataset are out-of-domain data.}\label{tbl1}
\begin{tabular*}{0.8\textwidth}{@{\extracolsep{\fill}}llcccc}
\toprule
& \multirow{2}{*}{\textbf{Dataset}} & \multirow{2}{*}{\textbf{Statistic}} & \multirow{2}{*}{\textbf{CN}} & \multicolumn{2}{c}{\textbf{Dementia}} \\
\cmidrule{5-6}
& & & & \textbf{AD} & \textbf{FTD}\\

\midrule
\multirow{4}{*}{In-domain} & \multirow{2}{*}{ADNI2} & No. subjects & \textbf{180} & \textbf{149} & \\
& & Age (Mean ± Std) & 73.4 ± 6.3 & 74.7 ± 8.1 & \\

\cmidrule{2-6}
& \multirow{2}{*}{NIFD} & No. subjects & \textbf{136} & & \textbf{150}\\
& & Age (Mean ± Std) & 63.5 ± 7.4 & & 63.9 ± 7.1\\

\midrule
\multirow{2}{*}{Out-of-domain}& \multirow{2}{*}{NACC} & No. subjects & 2182 & 485 & 37\\
& & Age (Mean ± Std) & 68.2 ± 10.9 & 72.3 ± 9.6 & 64.1 ± 6.9\\

\bottomrule
\end{tabular*}
\end{table}

\begin{figure}[ht]
\centering
\includegraphics[width=\textwidth]{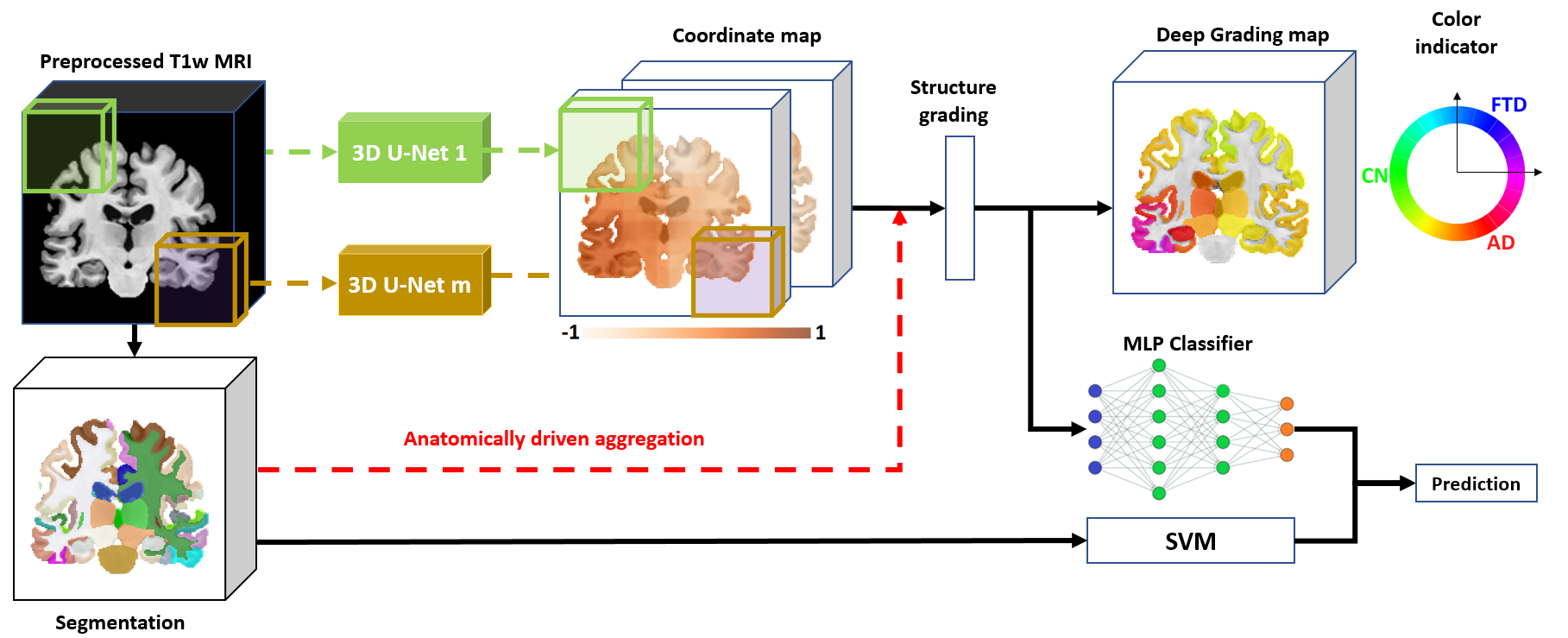}
\caption{An overview of the proposed multi-channel grading method. The T1w image, its segmentation and the deep grading map are taken from an AD patient.}
\label{figure:pipeline}
\end{figure}
\subsection{Preprocessing}
The preprocessing schema is composed of multiple steps: (1) denoising \cite{manjon_adaptive_2010}, (2) inhomogeneity correction \cite{tustison_n4itk_2010}, (3) affine registration into the MNI152 space ($181\times217\times181$ voxels at $1mm\times1mm\times1mm$) \cite{avants_reproducible_2011}, (4) intensity standardization \cite{manjon_robust_2008} and (5) intracranial cavity (ICC) extraction \cite{manjon_nonlocal_2014}. After preprocessing, we use AssemblyNet~\footnote{\url{https://github.com/volBrain/AssemblyNet}}  \cite{coupe_assemblynet_2020} to segment $s=133$ brain structures (see Figure \ref{figure:pipeline}). In this study, brain structure segmentation is used for two purposes. The first one is to aggregate structure grading for visualization and for the fully-connected classifier. And the second one is to compute the structure volumes (\ie normalized volume in \% of ICC) for classification using an SVM classifier (see Section \ref{section:method}).
\section{Method}
\label{section:method}

\subsection{Method overview}
Figure \ref{figure:pipeline} provides an overview of our method. After the preprocessing pipeline, a T1w MRI is downscaled with a factor of 2 to the size of $91 \times 109 \times 91$ voxels. The resulting image is then used to extract $k^3$ (\ie $k=5$) overlapping sub-volumes of the same size $32 \times 48 \times 32$ voxels and evenly distributed along the 3 image dimensions. We use $m=k^3$ (\ie $m=125$) U-Nets to grade these $m$ sub-volumes. The output of one U-Net has a size of $2 \times 32 \times 48 \times 32$ voxels (as the disease status is presented by a 2D point, see Section \ref{section:grading_multi_dementia}). The $m$ outputs are then used to reconstruct a DC map of size $2 \times 91 \times 109 \times 91$ voxels. This 2-channels map is upscaled to the same spatial size as the original input. After that, we compute the averaged DC for each brain structure with the help of an AssemblyNet-based brain segmentation \cite{coupe_assemblynet_2020} (see Section \ref{section:grading_multi_dementia}). The structure DC can be either used as input of a MLP classifier for classification or transformed into a 3D grading map for visualization (see Figure \ref{figure:pipeline}). Moreover, the structure volumes are used as input for an SVM classifier. Finally, we ensemble the results of two classifiers to get the diagnosis prediction.
\subsection{Deep Grading-based classification}
\label{section:grading_multi_dementia}
In medical imaging applications, it is more beneficial to provide the regions affected by diseases rather than just a classification result. For AD detection, several grading frameworks have been proposed to capture anatomical alterations caused by the disease \cite{coupe_simultaneous_2012, tong_novel_2017, coupe_scoring_2012, hett_graph_2018, nguyen_deep_2021}. The objective is to compute a 3D grading map reflecting the disease severity at the voxel level. In \cite{coupe_simultaneous_2012}, the authors assigned a score to each voxel by estimating how similar the surrounding region is to the corresponding region in healthy individuals and AD patients. Similarly, Tong \ea proposed to grade a small set of discriminative voxels across the whole brain using a sparse coding approach \cite{tong_novel_2017}. They demonstrated that the grading feature was efficient for the early detection of AD. More recently, Nguyen \ea extended the grading process to deep learning and proposed Deep Grading (DG) as an accurate and interpretable tool for AD detection \cite{nguyen_deep_2021}. Here, we propose to extend the DG framework to the problem of multi-class diagnosis.

As current grading systems only consider one pathology, its severity may be described by a single score. When many diseases are taken into account, we need to jointly determine which disease is present and also its severity. In this case, using a single scalar is impossible. To this end, we propose to assign each available class to a point in a 2D plan. Concretely, on a circle with a radius of 1, we assign $(-1,0)$ to CN, $(-\frac{\sqrt{3}}{2},0.5)$ to AD and $(\frac{\sqrt{3}}{2},0.5)$ to FTD (see the color indicator circle in Figure \ref{figure:pipeline}). All voxels outside of ICC are set to $(0,0)$ as they are not related to any pathology. With this definition, a predicted point depicting the disease status can be every point on that circle. Thus, the grading map is not only able to show the severity of each disease but also the common patterns of AD and FTD. We denote this approach as DGMD meaning deep grading for multi-dementia.

Based on the new definition of ground truth, each of our $m=125$ U-Nets takes a 3D sub-volume and outputs a DC map with 2 values for each voxel. For instance, when AD-like anatomical patterns are detected in a part of the brain, the produced values in this area should be close to $(\frac{\sqrt{3}}{2},0.5)$.

After that, we compute the averaged DC point for each brain structure. The obtained features are denoted as structure DC. By doing this, the grading map is encoded into a 2D matrix of size $2 \times s$ where s is the number of brain structures. Finally, we use a fully connected classifier to perform classification.
\subsection{Atrophy-based classification}
Besides the structure DC features, brain atrophy patterns are also important to identify AD and FTD patients. To exploit the atrophy features, we train a support vector machine (SVM) to perform the same classification task using normalized brain structure volumes. The output of the SVM model is combined with the MLP model to make the final decision. The detail of training the SVM and the ensembling process is provided in Section \ref{section:implementation_details}.
\subsection{Implementation details}
\label{section:implementation_details}
In each iteration of 10-fold cross-validation (see also Figure \ref{fig:data_split_procedure} about the data split in annexes), we used 10 data folds $d_i$ where $i \in \{1,..., 10\}$ as follows.
First, $d_1,...,d_7$ were used for training/validation of the 125 U-Nets.
Then, $d_1,...,d_7$ were re-used for training the MLP (and SVM) classifier and $d_8$ for its validation. After that, we used $d_9$ for ensembling the MLP and the SVM model. Finally, the ensemble model was evaluated on $d_{10}$.

To train each 3D U-Net, the data ($d_1,...,d_7$) is split into 80\%/20\% for training/validation. The data was common for all of $m=125$ U-Nets. However, each time we train a new U-Net, this data was combined and re-shuffled before splitting into training/validation to exploit the maximum information possible from our limited data. The loss used during training was voxel-wise mean square error (MSE) with Adam optimizer, batch size of 16 and a learning rate of 3e-4. The first U-Net was trained from scratch and was stopped after 400 epochs without improvement in validation loss. The following U-Nets took advantage of transfer learning from a neighborhood U-Net (see \cite{coupe_assemblynet_2020} for details) and thus, converted more quickly, their number of epochs for early stopping was set to 100.

To alleviate the overfitting phenomenon while training, we applied the following data augmentation schema: First, we randomly translated a sub-volume by $t \in \{-1,0,1\}$ voxel in its 3 axes. Second, we adapted Mixup \cite{zhang_mixup_2018} for DGMD. Concretely, given 2 pairs \{input voxel intensity, target DC point\}: $\{I_1, (x_1, y_1)\}$, $\{I_2, (x_2, y_2)\}$ taken from 2 subjects with class DC target $(X_1, Y_1)$ and $(X_2, Y_2)$ \footnote{$(x_i, y_i)$ and $(X_i, Y_i)$ can be different when the voxel is outside of ICC, see Section \ref{section:grading_multi_dementia}}, the mixup with a coefficient $\alpha \in (0, 1)$ is calculated as follows:
\begin{ceqn}
    \begin{align*}
        \begin{cases}
          I_{mixup} = \alpha I_1 + (1 - \alpha) I_2\\
          \phi_1 = atan2(Y_1, X_1)\\
          \phi_2 = atan2(Y_2, X_2)\\
          \phi_{mixup} = \alpha \phi_1 + (1 - \alpha) \phi_2\\
          x_{mixup} = \cos{\phi_{mixup}} * [\alpha (x_1^2 + y_1^2) + (1-\alpha) (x_2^2 + y_2^2)]\\
          y_{mixup} = \sin{\phi_{mixup}} * [\alpha (x_1^2 + y_1^2) + (1-\alpha) (x_2^2 + y_2^2)]
        \end{cases}
    \end{align*}
\end{ceqn}

When training the MLP classifier, we used cross-entropy loss with Adam optimizer, batch size of 8 and learning rate of 0.0003.

For the SVM classifier, we applied a grid search of three kernels (linear, polynomial, and radial basis function) and 500 values of C in $[10^{-5}, 10^5]$ on the validation set for tuning hyper-parameters. During training, due to the class imbalance nature of the dataset, we used balanced weights (available in scikit-learn library \cite{scikit-learn}) to compensate for the problem. 

To ensemble the MLP and SVM classifier, we made their prediction on the $d_9$. After that, we found a coefficient in $[0,1]$ that maximizes the balanced accuracy of the linear combination of MLP and SVM probabilities. Finally, the ensemble model was evaluated on $d_{10}$.

\section{Experimental results}
In this section, the 125 U-Nets were used as a feature extractor for every classification task. After the 10-fold cross-validation, we obtained in total 10 models.

To estimate the model performance on in-domain data, we evaluated 10 ensemble models on their corresponding in-domain test fold. By doing this, each testing sample was evaluated by one model and has one final prediction. We then concatenated all the prediction of 10 folds and compute different metrics based on that prediction.

To estimate the model performance on out-of-domain data, we evaluated 10 ensemble models on the out-of-domain data and averaged the output of these 10 models to boost the model generalization. By doing this, each testing sample was evaluated by ten models and had one final prediction. We then computed different metrics based on that prediction.
\subsection{Ablation study for binary classification tasks}
Table \ref{table:mixing_for_binary_classification} describes our ablation study for different binary classification tasks. This is done by evaluating 4 tasks: dementia diagnosis (\ie AD and FTD \vs CN), AD diagnosis (\ie AD vs CN), FTD diagnosis (\ie FTD vs CN) and 2-class differential diagnosis (\ie AD vs FTD). When training our classifiers, the 10 folds are remaining the same but all subjects with irrelevant classes are removed for each classification task. The balanced accuracy is used to assess the model performance, other metrics are also provided in annexes.

\begin{table}[ht]
    \centering
    \caption{Ablation study of our method for binary classification tasks. We use the balanced accuracy (BACC) to assess the performance. We perform 10-fold cross validation on ADNI+NIFD dataset to estimate the in-domain performance (exp. 1, 2, 3). Additionally, we evaluate on NACC dataset to estimate the out-of-domain performance (exp. 4, 5, 6) by averaging the outputs of 10 trained models. The results are presented in \%. {\color{red} Red}: best result, {\color{blue} Blue}: second result.}
    \label{table:mixing_for_binary_classification}
    \begin{tabular}{c@{\hskip 1em}c@{\hskip 1em}l@{\hskip 1em}c@{\hskip 1em}c@{\hskip 1em}c@{\hskip 1em}c@{\hskip 1em}c}
        \hline
        No.   & Evaluation & Features  & \makecell{Dementia \\ diagnosis\\Dem. \vs CN} & \makecell{AD\\diagnosis\\AD \vs CN}         & \makecell{FTD\\diagnosis\\FTD \vs CN} & \makecell{Differential \\ diagnosis\\AD \vs FTD}\\
        \hline
          &   &  & $N=615$ & $N=465$ & $N=466$ & $N=299$\\
        1 & \multirow{3}{*}{In-domain} & Volumes & 85.3 & 82.3 & 86.6 & 81.3\\
        2 & & DC & {\color{blue} \textbf{86.3}} & {\color{blue} \textbf{87.1}} & {\color{red} \textbf{91.0}}  & {\color{red} \textbf{94.3}}\\
        3 & & Ensemble &  {\color{red} \textbf{87.5}} &  {\color{red} \textbf{87.5}} & {\color{blue} \textbf{90.7}} & {\color{blue} \textbf{91.0}}\\
        
        \hline
          &   &  & $N=1627$ & $N=1605$ & $N=1353$ & $N=296$\\
        4 & \multirow{3}{*}{Out-of-domain} & Volumes & {\color{blue} \textbf{86.6}} & {\color{blue} \textbf{86.7}} & 87.0 & {\color{red} \textbf{88.9}}\\
        5 & & DC & 86.1 & 83.2 & {\color{blue} \textbf{88.6}}  & 84.0\\
        6 & & Ensemble &  {\color{red} \textbf{86.9}} &  {\color{red} \textbf{86.8}} & {\color{red} \textbf{89.1}} & {\color{blue} \textbf{87.1}}\\
        \hline
    \end{tabular}
\end{table}

Based on the results, we observe higher balanced accuracy of grade features than volume features for in-domain evaluation (exp. 1 \vs 2; ADNI+NIFD datasets) in every binary classification task. However, when evaluating on out-of-domain data (NACC dataset), the volume features are better than grade features (exp. 4 \vs 5) in all tasks except FTD diagnosis. Since the ensembling of two models can improve the performance in most of the cases compared to a single model, both grade and volume features are crucial for our classifications. However, they might focus on different characteristics of data (\eg grade features are more sensitive with FTD and volume features are more sensitive with CN, see Section \ref{section:performance_multi_class}), making different rankings for in-domain and out-of-domain datasets.
\subsection{Performance for multi-disease classification}
\label{section:performance_multi_class}
Table \ref{table:multi_class_classification} shows the results obtained for the 3-class differential diagnosis (\ie AD \vs CN \vs FTD). Different metrics are used to estimate the model performance: accuracy (ACC), balanced accuracy (BACC), area under curve (AUC) and sensitivity for each class. We observe that the volume features with the SVM classifier provide high CN sensitivity compared to grade features with the MLP classifier for both in-domain and out-of-domain evaluation. Besides, the grade features with MLP classifier provide high FTD sensitivity compared to volume features with SVM classifier for both in-domain and out-of-domain evaluation. These properties are important for the multi-class classification. Consequently, the combination of grade and volume consistently shows the best or second results in various metrics for both in-domain and out-of-domain evaluation. In the following, the results of our ensemble framework is used to compare with state-of-the-art methods.

\begin{table}[ht]
    \centering
    \caption{Performance of different models for the multiple disease classification. We denote ACC for accuracy, BACC for balanced accuracy, AUC for area under curve and Sen. for sensitivity. We perform 10-fold cross validation on ADNI+NIFD dataset to estimate the in-domain performance (exp. 1, 2, 3). Additionally, we evaluate on NACC dataset to estimate the out-of-domain performance (exp. 4, 5, 6) by averaging the outputs of 10 trained models. The results are presented in \%. The best and second performances are respectively in {\color{red} red} and {\color{blue} blue}.}
    \label{table:multi_class_classification}
    \begin{tabular}{c@{\hskip 1em}c@{\hskip 1em}l@{\hskip 3em}c@{\hskip 1em}c@{\hskip 1em}c@{\hskip 1em}c@{\hskip 1em}c@{\hskip 1em}c}
        \hline
        No. & Evaluation & Features    & ACC         & BACC & AUC & CN Sen. & AD Sen. & FTD Sen.\\
        \hline
        1 & \multirow{3}{*}{In-domain} & Volumes           & 81.3 & 77.2 & 91.5 & {\color{red} \textbf{92.4}} & 68.5 & 70.7 \\
        2 & & DC       & {\color{blue} \textbf{85.4}} & {\color{blue} \textbf{84.6}} & {\color{red} \textbf{94.3}} & 87.3 & {\color{red} \textbf{84.6}} & {\color{red} \textbf{82.0}}\\
        3 & & Ensemble  & {\color{red} \textbf{86.0}} & {\color{red} \textbf{84.7}} & {\color{blue} \textbf{93.8}} & {\color{blue} \textbf{89.6}} & {\color{blue} \textbf{83.2}} & {\color{blue} \textbf{81.3}} \\
        
        \hline
        4 & \multirow{3}{*}{Out-of-domain} & Volumes & {\color{red} \textbf{87.9}} & {\color{blue} \textbf{79.9}} & {\color{blue} \textbf{91.2}} & {\color{red} \textbf{91.6}} & {\color{blue} \textbf{72.6}} & 75.7 \\
        5 & & DC        & 82.7 & 79.2 & 88.8 & 85.2 & 71.3 & {\color{red} \textbf{81.1}}\\
        6 & & Ensemble   & {\color{blue} \textbf{87.1}} & {\color{red} \textbf{81.6}} & {\color{red} \textbf{91.6}} & {\color{blue} \textbf{89.6}} & {\color{red} \textbf{76.9}} & {\color{blue} \textbf{78.4}}\\
        \hline
    \end{tabular}
\end{table}
\subsection{Comparison with state-of-the-art methods}
In this section, we compare our method with two other deep learning based methods. In the first method, Hu \ea used a ResNet-like architecture for classification based on the intensities of a whole MR image \cite{hu_deep_2021}. They then used a guided backpropagation based method to visualize the dominant regions of AD and FTD pathologies. In the second method, Ma \ea firstly extract structure volume and cortical thickness (Cth) features from an MR image \cite{ma_differential_2020}. They then trained a Generative Adversarial Network (GAN) using these features and added an additional class for the fake data. At the inference time, the probability of this class is discarded for the final decision. 

We retrained the method of Hu \ea with the official publicly available code \footnote{\url{https://github.com/BigBug-NJU/FTD_AD_transfer}}. In the case of the second method, we  re-implement it based on the associated paper. For a fair comparison, we use the same 10 folds to train 10 models of each method. To train each model, 7 folds were used for training, 2 folds for validation. The remaining data fold was used to assess the in-domain performance. Finally, we applied the same data preprocessing pipeline used in our proposed method for training the state-of-the-art methods mentioned.

Table \ref{table:sota_2_classes} shows the comparison of our method with state-of-the-art methods for different problems of binary classification. Balanced accuracy (BACC) is used to assess the model performance. Other metrics, such as accuracy and area under curve are provided in the annexes. Our method consistently achieves the best results across all tasks, both in-domain (exp. 1, 2, 3) and out-of-domain diagnosis (exp. 4, 5, 6). This indicates the superior performance and effectiveness of our approach. Furthermore, our method demonstrates robustness to domain shift, surpassing other methods. On average, the performance drop between out-of-domain and in-domain evaluations for our method is only 1.7\%. In comparison, \cite{ma_differential_2020} exhibits an average drop of 2.1\%, while \cite{hu_deep_2021} shows a substantial average drop of 9.3\%. Overall, our method demonstrated high performance on different tasks and datasets and is more robust to external validation than other methods, highlighting its generalization capacity on unseen data and, thus, in clinical practice.

\begin{table}[ht]
    \centering
    \caption{Comparison of our method with current state-of-the-art methods for binary classification tasks. Our reported performances are the average of 10 repetitions and presented in \%. {\color{red} Red}: best result, {\color{blue} Blue}: second best result. The balanced accuracy (BACC) is used to assess the model performance. We denote Dem. for dementia (AD and FTD), CNN for convolutional neural network, GAN for generative adversarial network and Cth for cortical thickness.}
    \label{table:sota_2_classes}
    \begin{tabular}{c@{\hskip 0.6em}c@{\hskip 0.6em}l@{\hskip 0.6em}c@{\hskip 0.6em}c@{\hskip 0.6em}c@{\hskip 0.6em}c}
        \hline
        No. & Evaluation & Method  & \makecell{Dementia \\ diagnosis\\Dem. \vs CN} & \makecell{AD\\ diagnosis\\AD \vs CN}         & \makecell{FTD\\ diagnosis\\FTD \vs CN} & \makecell{Differential \\ diagnosis\\AD \vs FTD}\\
        \hline
        1 & \multirow{3}{*}{In-domain} & CNN on intensities \cite{hu_deep_2021} & 81.8 & 75.9 & 83.8 & {\color{blue} \textbf{82.3}}\\
        2 & & GAN on Cth and volumes \cite{ma_differential_2020} & {\color{blue} \textbf{85.1}} & {\color{blue} \textbf{85.3}} & {\color{blue} \textbf{85.7}} & 77.9\\
        3 & & Our method & {\color{red} \textbf{87.5}} & {\color{red} \textbf{87.5}}       & {\color{red} \textbf{90.7}} & {\color{red} \textbf{91.0}}\\
        \hline
        4 & \multirow{3}{*}{Out-of-domain} & CNN on intensities \cite{hu_deep_2021} & {\color{blue} \textbf{81.3}} & 76.1 & 68.0 & 61.2\\
        5 & & GAN on Cth and volumes \cite{ma_differential_2020} & 77.9 & {\color{blue} \textbf{86.6}} & {\color{blue} \textbf{80.8}} & {\color{blue} \textbf{80.5}}\\
        6 & & Our method & {\color{red} \textbf{86.9}} & {\color{red} \textbf{86.8}}       & {\color{red} \textbf{89.1}} & {\color{red} \textbf{87.1}}\\
        \hline
    \end{tabular}
\end{table}

Table \ref{table:sota_3_classes} presents the comparison of our method with the state-of-the-art methods under different metrics: accuracy (ACC), balanced accuracy (BACC), area under curve (AUC) and the sensitivity for each class (\ie CN, AD and FTD). Our method presents higher performance than other methods in global performance metrics (\ie ACC, BACC and AUC) for both in-domain and out-of-domain evaluation. Furthermore, our method presents similar performances in all ACC, BACC, AUC metrics, between in-domain and out-of-domain evaluations. This property is not observed in other methods~\cite{ma_differential_2020, hu_deep_2021}. It shows the high generalization capacity of our framework. In terms of sensitivity, our method achieves most of the time first or second place for all classes (\ie CN, AD and FTD).

\begin{table}[ht]
    \centering
    \caption{Comparison of our method with current state-of-the-art methods for 3-class differential diagnosis AD \vs FTD \vs CN. {\color{red} Red}: best result, {\color{blue} Blue}: second best result. We denote ACC for accuracy, BACC for balanced accuracy, AUC for area under curve, Sen. for sensitivity, CNN for convolutional neural network, GAN for generative adversarial network and Cth for cortical thickness.}
    \label{table:sota_3_classes}
    \begin{tabular}{@{\hskip 1em}c@{\hskip 1em}cl@{\hskip 3em}c@{\hskip 1em}c@{\hskip 1em}c@{\hskip 1em}c@{\hskip 1em}c@{\hskip 1em}c}
        \hline
        No. & Evaluation & Method    & ACC         & BACC & AUC & CN Sen. & AD Sen. & FTD Sen. \\
        \hline
        1 & \multirow{3}{*}{In-domain} & CNN on intensities \cite{hu_deep_2021} & 76.3 & 72.5 & {\color{blue} \textbf{90.0}} & 58.4 & {\color{red} \textbf{86.4}} & {\color{red} \textbf{96.5}}\\
        2 & & GAN on Cth and volume \cite{ma_differential_2020}      & {\color{blue} \textbf{77.1}} & {\color{blue} \textbf{75.9}} & 86.4 & {\color{blue} \textbf{80.4}} & 81.2 & 66.0\\
        3 & & Our method & {\color{red} \textbf{86.0}} & {\color{red} \textbf{84.7}} & {\color{red} \textbf{93.8}} & {\color{red} \textbf{89.6}} & {\color{blue} \textbf{83.2}} & {\color{blue} \textbf{81.3}}\\
        \hline
        4 & \multirow{3}{*}{Out-of-domain} & CNN on intensities \cite{hu_deep_2021} & {\color{blue} \textbf{85.2}} & 68.8 & 86.5 & {\color{blue} \textbf{68.0}} & {\color{red} \textbf{94.1}} & 48.6\\
        5 & & GAN on Cth and volume \cite{ma_differential_2020}      & 69.1 & {\color{blue} \textbf{74.6}} & {\color{blue} \textbf{87.5}} & 66.1 & {\color{blue} \textbf{82.1}} & {\color{blue} \textbf{75.7}} \\
        6 & & Our method & {\color{red} \textbf{87.1}} & {\color{red} \textbf{81.6}} & {\color{red} \textbf{91.6}} & {\color{red} \textbf{89.6}} & 76.9 & {\color{red} \textbf{78.4}}\\
        \hline
    \end{tabular}
\end{table}

Overall, our framework exhibits high performance and generalization capacity across various tasks, including binary and multi-disease diagnosis. However, it is important to note that there is a trade-off associated. Our framework, consisting of 125 U-Nets and an MLP classifier, comprises 393 million parameters, requires 25.9 TFLOPs for computation, takes 110 hours for training and has an inference time of 1.63 seconds (mainly due to the patch extracting and image reconstructing times). In comparison, the method of \cite{hu_deep_2021} presents 46 million parameters, 1 TFLOPs, 6 hours for training and an inference time of 1.4 $\times 10^{-3}$ seconds, while the method of \cite{ma_differential_2020} presents 0.11 million parameters, 6.8 $\times 10^{-6}$ TFLOPs, 0.4 hours for training and an inference time of 0.4 $\times 10^{-3}$ seconds.

\subsection{Interpretation of deep grading map}
\begin{figure}[ht]
\centering
\includegraphics[width=0.7\textwidth]{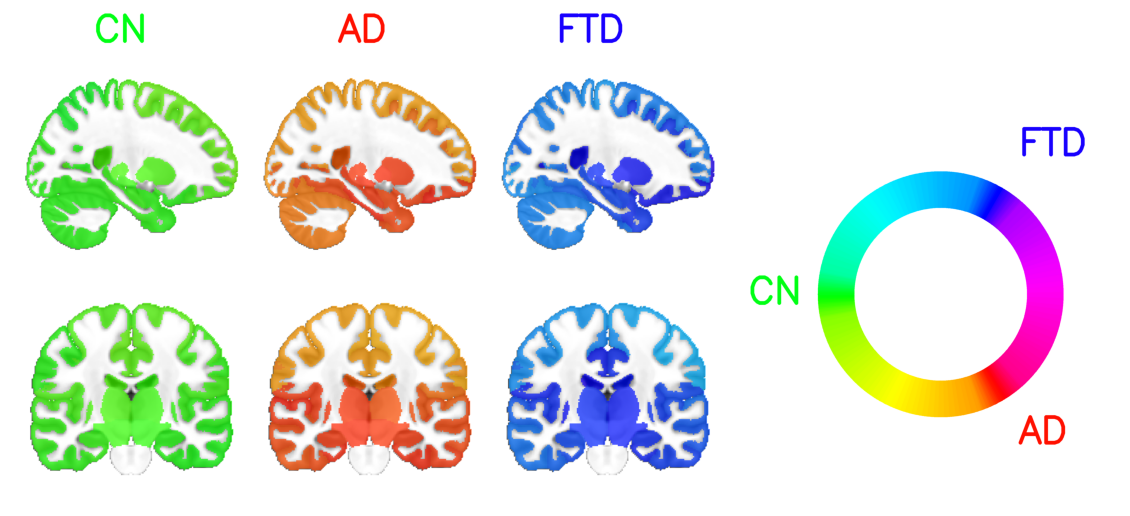}
\caption{Average grading map per group of subjects in the MNI152 space with neurological
orientation (with the right of the patient at the right).}
\label{figure:avg_map}
\end{figure}

To assess the interpretability provided by the grading map, we compute the averaged DC points (133 points for 133 brain structures) over subjects from each class. The considered subjects are taken from in-domain dataset. The averaged DC maps are transformed into grading maps for visualization. Figure \ref{figure:avg_map} shows sagittal and coronal views of these grading maps.

\begin{figure}[ht]
\centering
\includegraphics[width=0.45\textwidth]{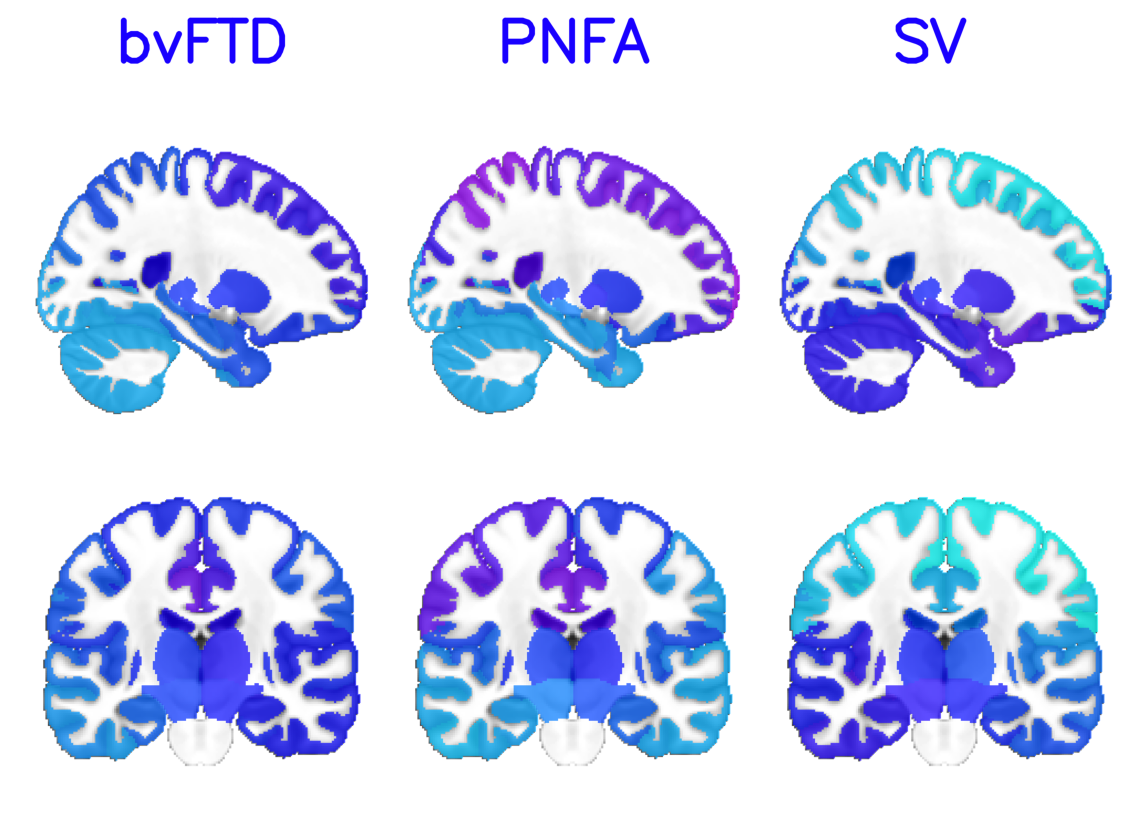}
\caption{Average grading map per variant of FTD in the MNI152 space with neurological
orientation (with the right of the patient at the right).}
\label{figure:avg_map_FTD}
\end{figure}

First, we can observe that our framework produces average grading maps well-separated for each class. As expected for the group of healthy people (\ie CN), all regions are detected as normal. For AD patients, the regions around the hippocampus are detected as AD-related patterns (red color). More generally, the temporal lobe is detected as strongly related to AD-like patterns in this population. The prevalence of AD in this region is widely documented \cite{schuff_mri_2009}. For the FTD class, we observe that FTD-like anatomical patterns are detected in similar areas. These results indicated that our method found diseases-specific anatomical anomalies (dissimilar patterns between AD and FTD) in similar locations for AD and FTD. This experiment highlights the need of grading map based on the multi-channel disease's coordinates. To further analyze our grading map, we compute the averaged map for each of its variants (\ie bvFTD, PNFA, SV) (see Figure \ref{figure:avg_map_FTD}).

We observe that the three sub-types present different FTD-related patterns. In the bvFTD group, the grading map highlights the frontal and temporal areas which are shown to be related to this pathology \cite{whitwell_distinct_2009}. In the PNFA group, the left frontal region \cite{boeve_advances_2022} and especially the left inferior frontal gyrus \cite{johns_dementia_2014} are highlighted which is typical of this syndrome. For the SV group, the left temporal pole is the most affected brain region. Indeed, this area presents typical atrophy in SV patients \cite{johns_dementia_2014}. We remark with the 3 variants of FTD that the disease severity is asymmetric, which is in line with the finding of Boeve \ea \cite{boeve_advances_2022}.

\begin{figure}[ht]
\centering
\includegraphics[width=\textwidth]{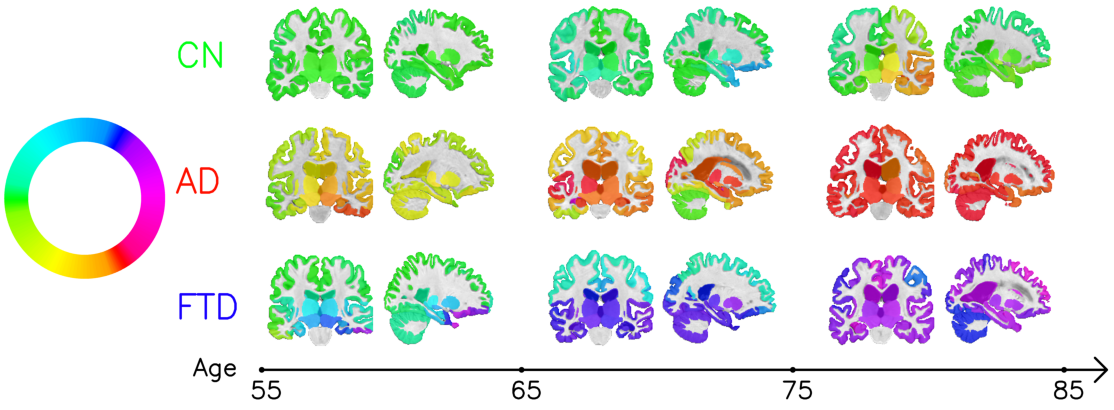}
\caption{Individual grading maps of each group of subjects with respect to age.}
\label{figure:individual_lifespan}
\end{figure}

Finally, we select typical deep grading maps of each class (\ie CN, AD and FTD) at different ages (see Figure \ref{figure:individual_lifespan}). We observe that in older healthy people, some areas have similar deep grading patterns with FTD \cite{chow_overlap_2008} and AD \cite{toepper_dissociating_2017}. In AD and FTD patients, both diseases start at a specific region (around the hippocampus for AD and frontotemporal lobes for FTD) and tend to expand to the whole brain over time.

\section{Discussion}
In this paper, we proposed a novel deep grading framework dedicated to binary and multi-disease classification problems. Moreover, we aimed at expanding our knowledge on AD and FTD disease-related patterns. So, beyond the predicted class for each individual, we provided also the color map indicating regions with specific disease patterns. The regions highlighted in each group of people (\ie CN, AD, bvFTD, PNFA, SV) as well as the asymmetric characterization provided by our framework are coherent with current knowledge of these diseases in the literature. Finally, we further investigate the three variants of FTD to describe the variability of this disease as suggested by Hu \ea \cite{hu_deep_2021}. This is expected to help clinicians to deeper understand FTD and to make more accurate diagnoses.

In this study, we take advantage of two types of biomarkers: structure grading and structure atrophy. While structure grading features provided by several U-Nets might offer information about anatomical patterns similarity with each class (\ie CN, AD, FTD), structure atrophy offers information about the abnormality of each brain structure in terms of size. Table \ref{table:multi_class_classification} demonstrates that the first biomarker can help to better detect FTD patients and the second one can accurately identify healthy people (\ie CN). As a result, our ensemble model improves the model performance not only in multi-disease tasks but also in many binary classification tasks (see Table \ref{table:mixing_for_binary_classification}).

This study is one among a few studies addressing the problem of multi-disease classification using sMRI data~\cite{bron_multiparametric_2017, kim_machine_2019, ma_differential_2020, hu_deep_2021}. We tried our best effort to make a fair comparison with state-of-the-art methods. Compared to these approaches, our method shows promising performance on different classification tasks (\ie dementia \vs CN, AD \vs CN, FTD \vs CN, AD \vs FTD and CN \vs AD \vs FTD). Experimental results demonstrate that our method is not only good with in-domain dataset but the learned patterns are generalizable, expressed by the lower drop of performance when evaluating on an out-of-domain dataset compared to other state-of-the-art methods. This characterization is shown in both binary classification and multi-disease classification tasks. This is very important in clinical practice where data are heterogeneous.

It is noteworthy that that we utilized the same (preprocessed) data to train our method and the state-of-the-art methods we compared to in this study. This choice was made based on the observation that the performance achieved by these methods using these preprocessed data was better than that achieved using raw data. Therefore, our preprocessing pipeline played a crucial role in enhancing the in-domain performance of both our method and the state-of-the-art methods while also contributing to improved generalization capacity on out-of-domain data.

Besides, the training data is an important factor leading to a good classification model. For instance, we used data coming from two different datasets with different classes: ADNI contains CN and AD patients while NIFD contains CN and FTD patients. These two datasets are chosen for their popularity and a lack of datasets with sufficient subjects for each class: CN, AD and FTD. However, it may exist some dataset-side biases. To alleviate the problem, only 3~Tesla images are selected as in \cite{hu_deep_2021}. It is possible that some people are misdiagnosed in these databases, where biological biomarkers are not always available, making a noisy ground-truth. Future works should consider the outlier removal to further improve model reliability. Finally, this study relies only on the sMRI at the baseline with the goal to detect brain diseases as early as possible. However, the patient's condition changes over time, it could be beneficial to use longitudinal data to make more accurate predictions and further track the progression of the disease.

\section{Conclusion}
In this paper, we propose a new framework for both disease detection (\ie AD + FTD \vs CN, AD \vs CN, FTD \vs CN) and differential diagnosis (\ie AD \vs FTD and CN \vs AD \vs FTD). First, we generate grading maps offering a meaningful visualization of the disease-related patterns. The grading scores can also be classified using a simple fully-connected classifier. Second, we propose to combine the obtained results with a support vector machine model using brain structure volumes to improve the model performance. By combining two types of features (\ie structure grading and structure atrophy), our method shows state-of-the-art performance in both disease detection and differential diagnosis.

\section*{Acknowledgement}
This work benefited from the support of the project DeepvolBrain of the French National Research Agency (ANR-18-CE45-0013). This study was achieved within the context of the Laboratory of Excellence TRAIL ANR-10-LABX-57 for the BigDataBrain project. Moreover, we thank the Investments for the future Program IdEx Bordeaux (ANR-10-IDEX-03-02 and RRI "IMPACT"), the French Ministry of Education and Research, and the CNRS for DeepMultiBrain project.

Datasets ADNI1 used for this project was funded by the Alzheimer's Disease Neuroimaging Initiative (ADNI) (National Institutes of Health Grant U01 AG024904) and DOD ADNI (Department of Defense award number W81XWH-12-2-0012). ADNI is funded by the National Institute on Aging, the National Institute of Biomedical Imaging and Bioengineering, and through generous contributions from the following: AbbVie, Alzheimer’s Association; Alzheimer’s Drug Discovery Foundation; Araclon Biotech; BioClinica, Inc.; Biogen; Bristol-Myers Squibb Company; CereSpir, Inc.; Cogstate; Eisai Inc.; Elan Pharmaceuticals, Inc.; Eli Lilly and Company; EuroImmun; F. Hoffmann-La Roche Ltd and its affiliated company Genentech, Inc.; Fujirebio; GE Healthcare; IXICO Ltd.; Janssen Alzheimer Immunotherapy Research \& Development, LLC.; Johnson \& Johnson Pharmaceutical Research \& Development LLC.; Lumosity; Lundbeck; Merck \& Co., Inc.; Meso Scale Diagnostics, LLC.; NeuroRx Research; Neurotrack Technologies; Novartis Pharmaceuticals Corporation; Pfizer Inc.; Piramal Imaging; Servier; Takeda Pharmaceutical Company; and Transition Therapeutics. The Canadian Institutes of Health Research is providing funds to support ADNI clinical sites in Canada. Private sector contributions are facilitated by the Foundation for the National Institutes of Health (www.fnih.org). The grantee organization is the Northern California Institute for Research and Education, and the study is coordinated by the Alzheimer’s Therapeutic Research Institute at the University of Southern California. ADNI data are disseminated by the Laboratory for Neuro Imaging at the University of Southern California.

TLDNI was funded through the National Institute of Aging, and started in 2010. The primary goals of FTLDNI were to identify neuroimaging modalities and methods of analysis for tracking frontotemporal lobar degeneration (FTLD) and to assess the value of imaging versus other biomarkers in diagnostic roles. The Principal Investigator of NIFD was Dr. Howard Rosen, MD at the University of California, San Francisco. The data are the result of collaborative efforts at three sites in North America. For up-to-date information on participation and protocol, please visit \url{http://memory.ucsf.edu/research/studies/nifd}. Data collection and sharing for this project was funded by the Frontotemporal Lobar Degeneration Neuroimaging Initiative (National Institutes of Health Grant R01 AG032306). The study is coordinated through the University of California, San Francisco, Memory and Aging Center. FTLDNI data are disseminated by the Laboratory for Neuro Imaging at the University of Southern California.

The NACC database is funded by NIA/NIH Grant U24 AG072122. NACC data are contributed by the NIA-funded ADRCs: P30 AG062429 (PI James Brewer, MD, PhD), P30 AG066468 (PI Oscar Lopez, MD), P30 AG062421 (PI Bradley Hyman, MD, PhD), P30 AG066509 (PI Thomas Grabowski, MD), P30 AG066514 (PI Mary Sano, PhD), P30 AG066530 (PI Helena Chui, MD), P30 AG066507 (PI Marilyn Albert, PhD), P30 AG066444 (PI John Morris, MD), P30 AG066518 (PI Jeffrey Kaye, MD), P30 AG066512 (PI Thomas Wisniewski, MD), P30 AG066462 (PI Scott Small, MD), P30 AG072979 (PI David Wolk, MD), P30 AG072972 (PI Charles DeCarli, MD), P30 AG072976 (PI Andrew Saykin, PsyD), P30 AG072975 (PI David Bennett, MD), P30 AG072978 (PI Neil Kowall, MD), P30 AG072977 (PI Robert Vassar, PhD), P30 AG066519 (PI Frank LaFerla, PhD), P30 AG062677 (PI Ronald Petersen, MD, PhD), P30 AG079280 (PI Eric Reiman, MD), P30 AG062422 (PI Gil Rabinovici, MD), P30 AG066511 (PI Allan Levey, MD, PhD), P30 AG072946 (PI Linda Van Eldik, PhD), P30 AG062715 (PI Sanjay Asthana, MD, FRCP), P30 AG072973 (PI Russell Swerdlow, MD), P30 AG066506 (PI Todd Golde, MD, PhD), P30 AG066508 (PI Stephen Strittmatter, MD, PhD), P30 AG066515 (PI Victor Henderson, MD, MS), P30 AG072947 (PI Suzanne Craft, PhD), P30 AG072931 (PI Henry Paulson, MD, PhD), P30 AG066546 (PI Sudha Seshadri, MD), P20 AG068024 (PI Erik Roberson, MD, PhD), P20 AG068053 (PI Justin Miller, PhD), P20 AG068077 (PI Gary Rosenberg, MD), P20 AG068082 (PI Angela Jefferson, PhD), P30 AG072958 (PI Heather Whitson, MD), P30 AG072959 (PI James Leverenz, MD).

\bibliographystyle{bibstyles/model1-num-names}
\bibliography{main}

\pagebreak
\clearpage
\appendix
\appendixpage

\section*{Our data splitting procedure}
For each cross-validation iteration, we used seven folds as training/validation data for our 125 U-Nets in the first stage. In the second stage, we reused this data as training data for our MLP and SVM classifiers. We took one more data fold as validation data for these classifiers. Once the MLP and the SVM were trained, one more data fold was used to find the coefficient to ensemble the two classifiers. Finally, we obtained an ensemble model of MLP and SVM and one remaining unused test fold.
\begin{figure}[h]
    \centering
    \includegraphics[width=0.8\textwidth]{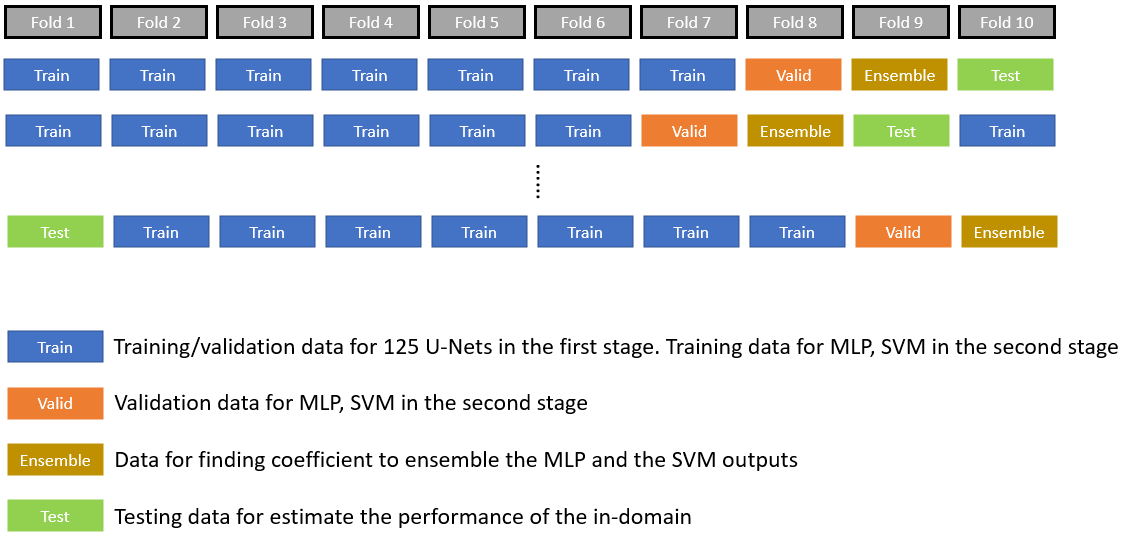}
    \caption{Our data split procedure}
    \label{fig:data_split_procedure}
\end{figure}

\section*{Ablation study using different metrics}
\vspace{1em}

\noindent
Ablation study of our method for binary classification tasks. We use the accuracy (ACC) to assess the performance. We perform 10-fold cross validation on ADNI+NIFD dataset to estimate the in-domain performance (exp. 1, 2, 3). Additionally, we evaluate on NACC dataset to estimate the out-of-domain performance (exp. 4, 5, 6) by averaging the outputs of 10 trained models. The results are presented in \%. {\color{red} Red}: best result, {\color{blue} Blue}: second best result.

\begin{center}
    \begin{tabular}{c@{\hskip 1em}c@{\hskip 1em}l@{\hskip 1em}c@{\hskip 1em}c@{\hskip 1em}c@{\hskip 1em}c@{\hskip 1em}c}
        \hline
        No.   & Evaluation & Features  & \makecell{Dementia \\ diagnosis\\Dem. \vs CN} & \makecell{AD\\diagnosis\\AD \vs CN}         & \makecell{FTD\\diagnosis\\FTD \vs CN} & \makecell{Differential \\ diagnosis\\AD \vs FTD}\\
        \hline
          &   &  & $N=615$ & $N=465$ & $N=466$ & $N=299$\\
        1 & \multirow{3}{*}{In-domain} & Volumes & 85.4 & 84.7 & 90.1 & 80.6\\
        2 & & Grades & {\color{blue} \textbf{86.3}} & {\color{blue} \textbf{87.5}} & {\color{blue} \textbf{93.1}}  & {\color{red} \textbf{94.6}}\\
        3 & & Ensemble &  {\color{red} \textbf{87.6}} &  {\color{red} \textbf{89.9}} & {\color{red} \textbf{93.7}} & {\color{blue} \textbf{93.3}}\\
        
        \hline
          &   &  & $N=1627$ & $N=1605$ & $N=1353$ & $N=296$\\
        4 & \multirow{3}{*}{Out-of-domain} & Volumes & {\color{red} \textbf{89.7}} & {\color{red} \textbf{92.2}} & {\color{blue} \textbf{96.6}} & {\color{red} \textbf{86.1}}\\
        5 & & Grades & 87.4 & 88.0 & 96.5  & 80.7\\
        6 & & Ensemble &  {\color{blue} \textbf{89.5}} &  {\color{blue} \textbf{91.2}} & {\color{red} \textbf{98.2}} & {\color{blue} \textbf{85.1}}\\
        \hline
    \end{tabular}
\end{center}

\vspace{3em}
\noindent
Ablation study of our method for binary classification tasks. We use the area under curve (AUC) to assess the performance. We perform 10-fold cross validation on ADNI+NIFD dataset to estimate the in-domain performance (exp. 1, 2, 3). Additionally, we evaluate on NACC dataset to estimate the out-of-domain performance (exp. 4, 5, 6) by averaging the outputs of 10 trained models. The results are presented in \%. {\color{red} Red}: best result, {\color{blue} Blue}: second best result.

\begin{center}
    \begin{tabular}{c@{\hskip 1em}c@{\hskip 1em}l@{\hskip 1em}c@{\hskip 1em}c@{\hskip 1em}c@{\hskip 1em}c@{\hskip 1em}c}
        \hline
        No.   & Evaluation & Features  & \makecell{Dementia \\ diagnosis\\Dem. \vs CN} & \makecell{AD\\diagnosis\\AD \vs CN}         & \makecell{FTD\\diagnosis\\FTD \vs CN} & \makecell{Differential \\ diagnosis\\AD \vs FTD}\\
        \hline
          &   &  & $N=615$ & $N=465$ & $N=466$ & $N=299$\\
        1 & \multirow{3}{*}{In-domain} & Volumes & {\color{blue} \textbf{92.3}} & 91.3 & 93.7 & 87.6\\
        2 & & Grades & 92.1 & {\color{blue} \textbf{93.1}} & {\color{red} \textbf{96.2}}  & {\color{red} \textbf{99.2}}\\
        3 & & Ensemble &  {\color{red} \textbf{93.5}} &  {\color{red} \textbf{93.9}} & {\color{blue} \textbf{95.3}} & {\color{blue} \textbf{96.6}}\\
        
        \hline
          &   &  & $N=1627$ & $N=1605$ & $N=1353$ & $N=296$\\
        4 & \multirow{3}{*}{Out-of-domain} & Volumes & {\color{blue} \textbf{95.6}} & {\color{blue} \textbf{95.5}} & {\color{blue} \textbf{98.6}} & {\color{red} \textbf{94.1}}\\
        5 & & Grades & 94.2 & 93.4 & 92.3  & 87.3\\
        6 & & Ensemble &  {\color{red} \textbf{96.0}} &  {\color{red} \textbf{95.9}} & {\color{red} \textbf{99.2}} & {\color{blue} \textbf{92.7}}\\
        \hline
    \end{tabular}
\end{center}
\section*{Comparison with current state-of-the-art methods using different metrics}
\vspace{1em}
\noindent
Comparison of our method with current state-of-the-art methods for binary classification tasks. Our reported performances are the average of 10 repetitions and presented in \%. {\color{red} Red}: best result, {\color{blue} Blue}: second best result. The accuracy (ACC) is used to assess the model performance.

\begin{center}
    \begin{tabular}{c@{\hskip 0.6em}c@{\hskip 0.6em}l@{\hskip 0.6em}c@{\hskip 0.6em}c@{\hskip 0.6em}c@{\hskip 0.6em}c}
        \hline
        No. & Evaluation & Method  & \makecell{Dementia \\ diagnosis\\Dem. \vs CN} & \makecell{AD\\ diagnosis\\AD \vs CN}         & \makecell{FTD\\ diagnosis\\FTD \vs CN} & \makecell{Differential \\ diagnosis\\AD \vs FTD}\\
        \hline
        1 & \multirow{3}{*}{In-domain} & CNN on intensities \cite{hu_deep_2021} & 82.0 & 81.7 & 87.3 & {\color{blue} \textbf{82.3}}\\
        2 & & GAN on Cth and volume \cite{ma_differential_2020} & {\color{red} \textbf{85.0}} & {\color{red} \textbf{87.3}} & {\color{red} \textbf{88.6}} & 77.9\\
        3 & & Our method & {\color{blue} \textbf{87.5}} & {\color{blue} \textbf{89.0}}       & {\color{blue} \textbf{93.3}} & {\color{red} \textbf{91.0}}\\
        \hline
        4 & \multirow{3}{*}{Out-of-domain} & CNN on intensities \cite{hu_deep_2021} & {\color{blue} \textbf{86.8}} & {\color{blue} \textbf{86.8}} & {\color{red} \textbf{97.3}} & {\color{red} \textbf{85.8}}\\
        5 & & GAN on Cth and volume \cite{ma_differential_2020} & 67.3 & 85.8 & 75.3 & 75.3\\
        6 & & Our method & {\color{red} \textbf{87.5}} & {\color{red} \textbf{90.0}}       & {\color{blue} \textbf{96.8}} & {\color{blue} \textbf{80.7}}\\
        \hline
    \end{tabular}
\end{center}

\vspace{3em}
\noindent
Comparison of our method with current state-of-the-art methods for binary classification tasks. Our reported performances are the average of 10 repetitions and presented in \%. {\color{red} Red}: best result, {\color{blue} Blue}: second best result. The area under curve (AUC) is used to assess the model performance.

\begin{center}
    \begin{tabular}{c@{\hskip 0.6em}c@{\hskip 0.6em}l@{\hskip 0.6em}c@{\hskip 0.6em}c@{\hskip 0.6em}c@{\hskip 0.6em}c}
        \hline
        No. & Evaluation & Method  & \makecell{Dementia \\ diagnosis\\Dem. \vs CN} & \makecell{AD\\ diagnosis\\AD \vs CN}         & \makecell{FTD\\ diagnosis\\FTD \vs CN} & \makecell{Differential \\ diagnosis\\AD \vs FTD}\\
        \hline
        1 & \multirow{3}{*}{In-domain} & CNN on intensities \cite{hu_deep_2021} & 88.5 & 86.1 & 89.3 & {\color{blue} \textbf{90.2}}\\
        2 & & GAN on Cth and volume \cite{ma_differential_2020} & {\color{blue} \textbf{91.1}} & {\color{red} \textbf{89.9}} & {\color{red} \textbf{91.2}} & 82.6\\
        3 & & Our method & {\color{red} \textbf{93.5}} & {\color{blue} \textbf{93.7}}       & {\color{blue} \textbf{95.0}} & {\color{red} \textbf{95.0}}\\
        \hline
        4 & \multirow{3}{*}{Out-of-domain} & CNN on intensities \cite{hu_deep_2021} & 88.4 & 88.9 & 86.2 & 75.3\\
        5 & & GAN on Cth and volume \cite{ma_differential_2020} & {\color{blue} \textbf{92.4}} & {\color{blue} \textbf{93.3}} & {\color{blue} \textbf{87.8}} & {\color{blue} \textbf{93.5}}\\
        6 & & Our method & {\color{red} \textbf{93.8}} & {\color{red} \textbf{94.4}}       & {\color{red} \textbf{90.3}} & {\color{red} \textbf{95.7}}\\
        \hline
    \end{tabular}
\end{center}

\end{document}